\documentclass[preprint,12pt]{elsarticle}

\usepackage{multirow}

\usepackage{natbib}
\usepackage{graphicx}
\usepackage{dcolumn}
\usepackage{bm}

\usepackage{url} 
\usepackage{lineno}
 \usepackage{amsthm}

\usepackage[utf8]{inputenc}
\usepackage[T1]{fontenc}
\usepackage{mathptmx}
\usepackage{etoolbox}
\usepackage{ulem} 
\usepackage{xcolor}
\usepackage{tablefootnote}

\newcommand{\rev}[1]{#1}

\makeatother
\begin{document}


\title{Distributed network of smartphone sensors: a new tool for scientific field measurements}

\author[PMMH]{Jishen Zhang}
\author[ISTerre]{Nicolas Mokus} 
\author[PMMH]{Jules Casoli}
\author[PMMH]{Antonin Eddi}
\author[PMMH]{Stéphane Perrard\corref{cor1}}
\cortext[cor1]{stephane.perrard@espci.fr}

\address[PMMH]{PMMH, CNRS, ESPCI Paris, Université PSL, Sorbonne Université, Université de Paris, F-75005 Paris, France}
\address[ISTerre]{ISTerre, Université Grenoble Alpes, Université Savoie Mont Blanc, CNRS, IRD, IFSTTAR, F-38058 Grenoble, France}


\date{\today}

\begin{abstract}
Smartphones sensors are now commonly used by a worldwide audience thanks to their availability, high connectivity, and versatility. Here, we present a methodology to use a collection of smartphones, namely a fleet, as a distributed network of time-synchronized mechanical sensors. We first present the mechanical tests we develop to evaluate the smartphone sensor accuracy. We then describe how to use efficiently a distributed network of smartphones as autonomous sensors. We use a combination of an Android application hosted on each phone (\texttt{Gobannos}), and a server application (Phonefleet) on a controlling host to perform the tasks in parallel remotely. We implement in particular a time synchronization protocol based on UDP communication. We achieved an accuracy of the smartphone clock synchronisation of 60 microseconds. Using two test cases in realistic outdoor conditions, we eventually prove the reliability of a smartphone fleet to measure  mechanical wave measurements in field conditions.
\end{abstract}

\maketitle

\section{Introduction}\label{sec:intro}

By November 2024, mobile subscriptions have reached 8.4 billion, including 7.14 billion for smartphones~\citep{ericsson2024ericsson}. Smartphones have become a key tool for educational and scientific research purposes thanks to their accessibility to the general public~\citep{gao2016smartphone,raju2024commentary} and to the versatility of their built-in micro-electro-mechanical system (MEMS) sensors. A typical MEMS is a highly integrated, silicon-based sensor system fabricated using semiconductor processes~\citep{maluf2002introduction}. Its low cost, compact size, lightweight design, and low power consumption make a MEMS ideal for integration into smartphones, drones, tablets, and wearable devices~\citep{daponte2013state}, enabling applications such as step counting and orientation tracking~\citep{bao2004activity, naqvib2012step,davidson2016survey}, gaming~\citep{hopfner2013measuring, kolakowska2020review}, or navigation~\citep{li2015real, mostafa2019novel}. 

In the context of physics education, following the COVID-19 pandemic, smartphones have gained popularity as flexible tools for providing innovative teaching methodologies, particularly at the university level~\cite{bobroff2020teaching,bouquet202061, o2021guide}. Numerous examples of tabletop physics experiments using smartphone as a measurement device can be found, such as speed of sound measurement using the embedded microphone~\cite{kasper2015stationary}; earth's rotation estimate using the accelerometer~\cite{vandermarliere2021detect}; and light polarization analyses using ambient light sensor~\cite{monteiro2017polarization}. A comprehensive reviews of applications can be found in Organtini~\cite{organtini2021physics} or in Kuhn $\&$ Vogt~ \cite{kuhn2022smartphones}. 

The versatility of smartphone MEMS sensors has also drawn significant attention from researchers across various scientific domains. The high connectivity and ubiquity of smartphones enable large-scale, real-time data collection through a geographically distributed network of interconnected devices. For public health research, investigations on the data collection and analysis of energy expenditure from smartphone users~\cite{pande2013energy} have shown that smartphone-embedded IMU (Inertial Measurement Unit) sensors such as accelerometers and gyroscopes, can accurately monitor the user's physical activities, thus enhance the precision of individual health assessments. Smartphone inertial sensors have also been widely used for human activity recognition, where relevant motion features span a broad range of time scales, from rapid impacts to slower postural changes. Recent studies show that multiscale signal representations significantly improve activity classification while remaining compatible with the limited computational resources of smartphones.~\cite{quan2025msa, tang2022multiscale} In a review by Lee et al.~\cite{lee2018review} on smartphone applications in geoscience research, the authors highlight the growing use of smartphones in tasks traditionally carried out with specialized tools, such as geological mapping, seismic activity detection, and natural hazard assessment.

However, field measurements often require multi-point data collection with sensors spatially distributed over a wide area while remaining synchronized in time. Examples of such applications include predicting solar magnetic storms and assessing their impact on the Earth's magnetosphere~\cite{odenwald2022can}, as well as performing modal analysis of civil engineering structures~\cite{gueguen2020slow}, where a large number of data points are crucial for reconstructing higher-order spatial modes~\cite{cunha2006experimental,rainieri2014operational}. Distributed sensor measurements in seismology using smartphones have been introduced via real-time data collections for earthquake detection using the application \texttt{MyShake}~\cite{kong2016myshake}. It allows accurate localization of the earthquake epicenters and detects seismic waves propagation in real time, contributing to early warning systems~\cite{reilly2013mobile,kong2016myshake}.
These applications do not rely on sophisticated, high-cost sensors but rather utilize frugal, cost-effective sensor networks, such as smartphone-based MEMS sensors fleets on a large scale, often in regions geographically inaccessible through traditional methods.


Several mobile applications (\texttt{PhyPhox, FizziQ})~\cite{staacks2018advanced,bilgin2022stem} have been developed for individual users to facilitate sensor communication and to display data in real-time. However, there is still a notable lack of a technical way to control multiple smartphones by a single user. In contrast to single smartphone measurement that can be performed with manual interactions, distributed measurements require an automated process involving remote communication for data sampling and collection, time synchronization between smartphones, and parallel task planification. Another key challenge lies in the optimization of human-machine interaction (HMI) for a single user to control a \rev{smartphone fleet}.


In this study, we present a methodology to perform simultaneous multi-sensor measurements using a fleet of identical smartphones. The article is structured as follows: we first review the measurement accuracy of a single smartphone using several test setups. We characterize in particular the ambient noise amplitude and spectrum of a single smartphone placed in a still environment. We then perform accelerometer and gyroscope sensor calibration on a turntable platform and we determine the sensor location within the phone~\citep{mau2016locating}. This retro-engineering approach allows to access information that are otherwise proprietary for most commercial products. Next, we develop an automated remote communication and control protocol for data recording across the smartphone fleet, achieving time synchronization with a typical error of \rev{60~$\mu$s}. To do so, we introduce a custom Android-based application, \texttt{Gobannos}~\cite{Gobannos_2025}, which integrates memory allocation for local data storage, remote communication, and time synchronization. We eventually provide two examples of mechanical wave measurements, demonstrating the feasibility of reliable data collection \rev{in field conditions} using a large fleet of smartphones.


\section{Smartphones as a multi-sensor fleet \label{sec:fleet}}

To create a fleet of autonomous, multi-physics sensors, we selected the Redmi 10A smartphone, balancing the cost and the quality of the embedded IMU sensors (accelerometer, gyroscope and magnetometer). \rev{We have also tested and benchmarked the sensors of the Fairphone 4. Still,} the methodology described in this manuscript may apply to all types of smartphones. The fleet is composed of 66 smartphones, numbered from 0 to 65. We use concomitantly the accelerometer, gyroscope, magnetometer and GPS sensors. 
To access and utilize the smartphone sensors, we initially employed the \texttt{Phyphox} application~\cite{staacks2018advanced}, developed at the University of Aachen. \texttt{Phyphox} offers an intuitive interface for acquiring data from individual sensors or from multiple sensors simultaneously, along with a standard URL communication protocol to retrieve data from the smartphone. We configured a module in \texttt{Phyphox} to simultaneously collect accelerometer, gyroscope, magnetometer, and GPS data at their respective maximum sampling frequencies. In a second step, we develop an Android application, named \texttt{Gobannos}~\cite{Gobannos_2025}, better adapted to parallelized, continuous and time synchronized acquisitions without physical intervention on the phones. 
This section is organized as follow. Section~\ref{sec:test} presents the mechanical tests performed on the smartphone sensors. Section~\ref{sec:fleetRC} details the architecture we develop to control a smartphone fleet from a single computer. \rev{Some of the additional tests we performed on the smartphone fleet are presented in appendices.}

\begin{figure}[t!]
  \includegraphics[width=11cm]{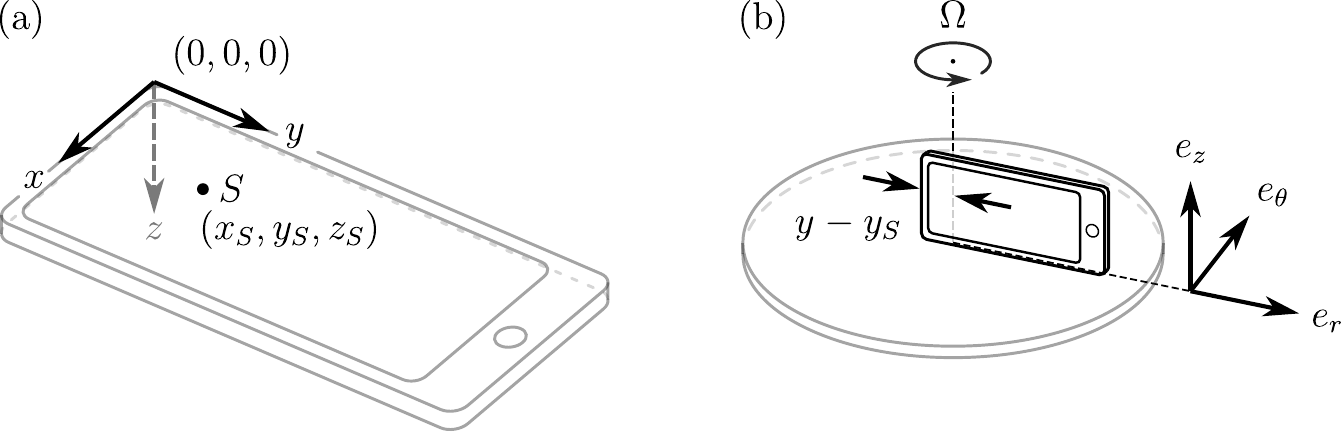}
  \caption{\label{fig:Sketch} (a) Sketch of the smartphone coordinate system ($0,x,y,z$) with the origin located on the top-right up corner of the phone's bounding box. The accelerometer sensor is located at point $S$. (b) Sketch of the turning-table setup used for the sensor tests.}
\end{figure}

\subsection{Sensor tests}\label{sec:test}

The coordinate system of the smartphone sensors is shown in Fig.~\ref{fig:Sketch}(a), according to the direction of the acceleration of gravity measured along the three main axes~(Fig.~\ref{fig:Sketch}(a)). Note that the $(x,y,z)$ coordinate frame of the phone forms an indirect basis. The smartphones are equipped with built-in IMU sensors model ICM-42670-P fabricated by TDK InvenSense~\cite{invenSense2021}. 

We first characterize the noise amplitude of the smartphone sensors using recordings in a quiescent environment. We then mount a smartphone on a rotating table to identify the exact location $S$ of the accelerometer. We eventually use the same experimental set up to calibrate the gyroscope sensor. 

\subsubsection{Noise amplitude and spectrum}\label{sec:noise}

\begin{figure}[t!]
  \includegraphics[width=\columnwidth]{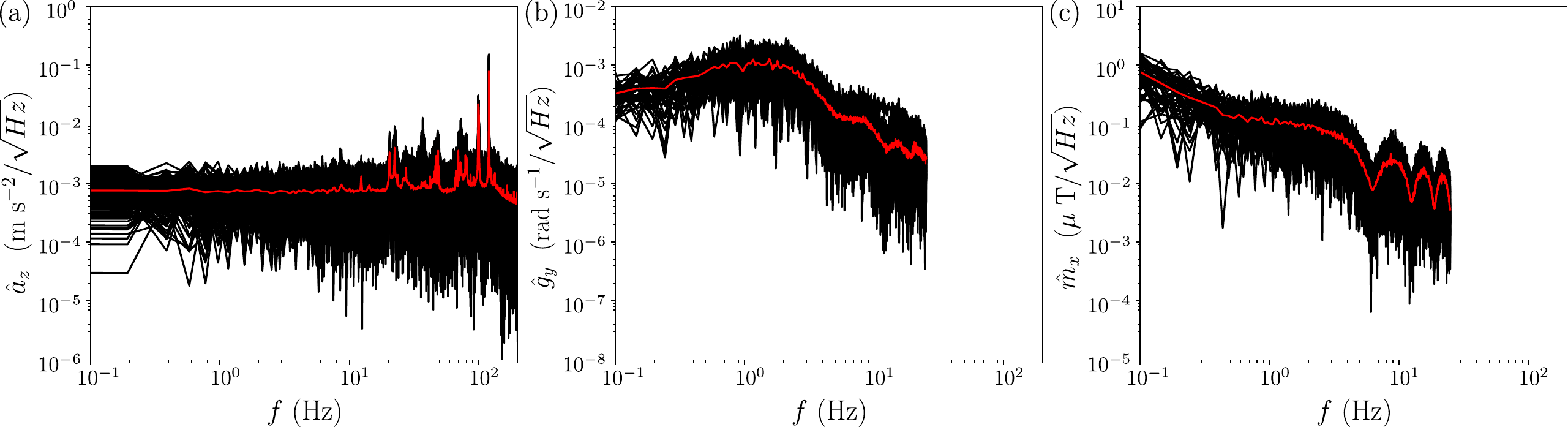}
  \caption{\label{fig:spectrum_calibration} Noise spectrum of accelerometer (a), gyroscope (b) and magnetometer (c) computed on a noise recordings of 5  minutes 
  (red curve). Each black line corresponds to a single phone recording. Note that the sampling frequency is around 400~Hz for the accelerometer, while it is only 50~Hz for both gyroscope and magnetometer. }
\end{figure}

We first perform tests on the noise level of the smartphone sensors. We perform five-minute recordings in a quiescent environment with the  accelerometer, the gyroscope and the magnetometer. The typical sampling frequency of the accelerometer is 400~Hz, with variations from a smartphone to another of about 1\%. Note that for each smartphone, the sampling frequency is stable over time. A typical value of the sampling frequency is $fs = 403.96 \pm 0.06$~Hz (smartphone \#~30). The typical acceleration error is $\sigma_a$ = 0.02 m s$^{-2}$ in the direction perpendicular to gravity and $\sigma_a$ = 0.07 m s$^{-2}$ along the direction of gravity. We notice that the gyroscope was sold by the constructor as a 400~Hz sampling frequency sensor, however in practice, it only operates at 50~Hz. We notice that the gyroscope measurements have indeed the same nominal sampling frequency of 400 Hz as the accelerometer. However, the gyroscope samples at 50 Hz, with repeated constant values added to artificially increase the frequency to 400 Hz. Typical error is $\sigma_g = 0.004 $ rad/s. We tested another smartphone model, the Fairphone 4, whose accelerometer also operate around 400Hz. The effective gyroscope sampling frequency in this case was equal to the expected value of 400Hz.
    
We interpolate the signal on a regular grid in time at a sampling frequency $f_s$ close to their maximum sensor sampling frequency (resp. 400~Hz for $a$ and 50~Hz for $g$ and $m$), and compute the temporal Fourier transform $\hat a_i$, $\hat g_i$ and $\hat m_i$ of the acceleration, angular velocity, and magnetic field components (resp. $a_i$, $g_i$ and $m_i$). \rev{To compute the Fourier transform, we use standard signal processing procedure, including zero padding of 4 times the signal on both ends, Hanning windowing and no overlap. The samples corresponds to a recording of 10s each}. The power spectrum of one component for each sensor type (resp. $a_z$, $g_y$ and $m_x$) is shown in Fig.~\ref{fig:spectrum_calibration}(a), (b) and (c) as a function of the frequency. \rev{The black curve corresponds to one Fourier transform, while the red line correspond to the average over 30 consecutive realizations.} The noise acceleration spectrum is flat in the entire range of tested frequency $f \in [0.1,200]$~Hz. We observe residual vibrations peaks at large frequencies. However, they may be specific to the contact between the phone and the ground, and the amplitude and locations of the peaks may not be reproducible. The noise spectrum of the gyroscope data is colored (see Fig.~\ref{fig:spectrum_calibration}(b)) with a maximum of sensitivity around 1 Hz. The magnetometer exhibits a typical error $\sigma_m = 1~\mu$T about one hundred time smaller than the Earth magnetic field. The associated noise spectrum is also colored, with a steeper slope than the gyroscope, and secondary lobes above 5~Hz. \rev{Note that the spectrum of the magnetometer exhibits zeros at $f_S/2$, $f_S/4$ and $f_S/8$, with $f_S = 50$Hz. These zeros are the signature of a moving average filter on 8 consecutive time steps, performed on the magnetometer data. As a consequence, the magnetometer signal is delayed by 4 time steps $4/f_S$ with respect to the other sensors. Using additionnal test on an oscillating pendulum in the context of the application example shown in Sec.~\ref{sec:chainPendule}, we have checked that the magnetometer data are indeed delayed by $4/f_S = 0.08$s with respect to the accelerometer and the gyroscope recordings.}

About 15\% of the smartphones showed a parasitic beat signal on the gyroscope sensors, and were relegated to the last number of the series. The first 50 smartphones sensors all show comparable quality of their sensors. Note that the smartphones do not share continuous serial numbers, and do not necessarily come from the same manufactured series. 

\subsubsection{Accelerometer sensor location}\label{sec:MEMSlcs}

\begin{figure}[t!]
  \includegraphics[width=\columnwidth]{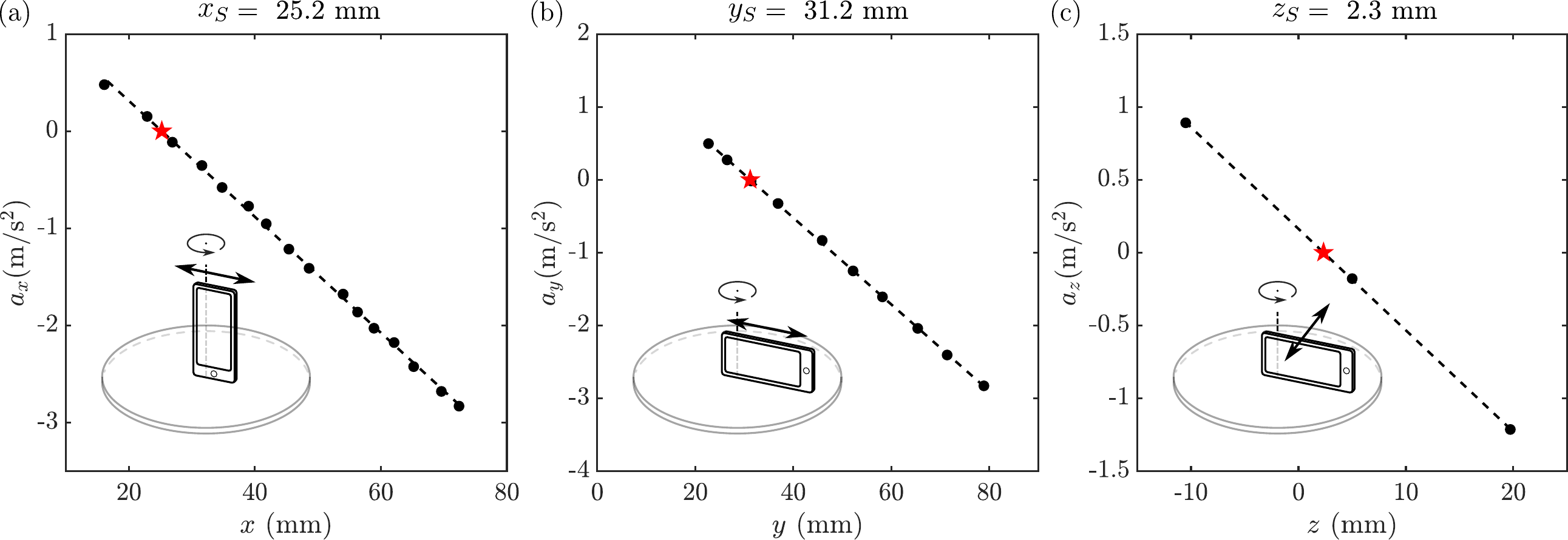}
  \caption{\label{fig:acceleroCentre} Determination of the accelerometer sensor location along the three main smartphone axis. a), b) and c) Measurements of respectively $a_x$, $a_y$, $a_z$ as a function of the sliding position $x$ (resp. $y$, $z$), for a rotation axis aligned with $y$ (resp. $x$ and $x$). In the three cases we observe a linear relationship, which intersects zero at the sensor position $x_S$ (resp. $y_S$, $z_S$) denoted by the red stars.}
\end{figure}

We mount a smartphone on a table, rotating around a vertical axis ($Z$) at an angular frequency $\Omega$ (Fig.~\ref{fig:Sketch}(b)). We conduct three series of experiments, with the smartphone placed with its axis $y$ (resp. $x$ and $x$) aligned with the vertical axis $e_z$. We vary the smartphone position with respect to the rotation axis by sliding it along the axis $x$, $y$ and $z$ respectively. The slide direction is either radial (sketches of Fig.~\ref{fig:acceleroCentre}(a) and (b)), or tangential (sketch of Fig.~\ref{fig:acceleroCentre}(c)). We then simultaneously record the accelerometer and the gyroscope signals for 60s. From a force balance in the rotating frame of reference, we expect the radial acceleration (along $e_r$) to be $a_x$ = $r \Omega^2$ (resp. $a_y$ and $a_z$). The acceleration $a_x$ vanishes when the sensor is located on the rotation axis. Note that origin of space is taken at the upper top right corner of the phone, as sketched on Fig.~\ref{fig:Sketch}(a). Fig.~\ref{fig:acceleroCentre}(a,b,c) show respectively the radial accelerations $a_x$, $a_y$ and $a_z$ for the smartphone rotating respectively along $y$, $x$ and $x$ axes at the angular frequency $\Omega = 7.7$~rad/s as a function of the smartphone locations. We indeed find a linear relationship with $r$, and we extract the position of the sensor along the three axes by interpolating the curves in $a=0$, represented by the black stars. We found $x_S = 25.2 \pm 0.1$~mm, $y_S = 31.2 \pm 0.1$~mm and $z_S = 2.3 \pm 0.1$~mm. The slope of the linear relation, $a_x$ = $r \Omega^2$ can be used to check the accuracy of the accelerometer, knowing the rotation rate $\Omega$. We found an excellent agreement (less than 1 \% error) in all three direction of orientations.

\subsubsection{Accelerometer $\&$ gyroscope calibration}\label{sec:MEMScalib}

\rev{The gyroscope accuracy has also been evaluated on the rotating table, using the accelerometer data as the reference. The results are detailed in~\ref{sec:gyrocalib}. For the Redmi 10A, we observe a significant error on the measured angular frequency, which depends on the distance between the sensor and the rotation axis. The gyroscope is a vibrating structure gyroscope (VSG) which is based on the Coriolis effect applied on a vibrating mass. As a consequence, in a non Galilean frame of reference, the gyroscope measures a combination of inertial acceleration and Coriolis force, and the sensor reading deviates from the true angular frequency, both in magnitude and in direction.}

 We conclude that in practice, the gyroscope of the redmi 10A is reliable only to measure pure rotational motions around the gyroscope sensor, and cannot be used in general in combination with the accelerometer to decompose arbitrary superposition of translational and rotational motions. However, in some limit cases, in particular when the center of rotation is known, we expect the gyroscope to be reliable. One may refer to Fig.~\ref{fig:calib_g_a}(f) to evaluate the systematic error as a function of both the angular frequency and the distance $r$ between the sensor and the axis of rotation.
 
 \rev{We have also conducted tests on a Fairphone 4, the results are shown in~\ref{sec:gyrocalib}. We found that the gyroscope was accurate up to the maximum tested angular frequency of 12rad/s. We deduce that the limitation we observed on the gyroscope sensor of the Redmi 10A are phone dependent, and each smartphone model must be tested before use, to benchmark the accuracy of their sensors in various configurations.}
 
\subsection{Smartphone fleet remote control}\label{sec:fleetRC}

\subsubsection{Remote control of a single smartphone}\label{sec:fleetRC1}

Numerous smartphone applications exist to record the phone sensor data. We first use the application \texttt{Phyphox}~\cite{staacks2018advanced} developed at the university of Aachen, which provides a user-friendly interface to acquire data from several sensors in parallel. \texttt{Phyphox} provides in particular a standard URL communication protocol to send instructions (START, STOP, CLEAR) from a distant host. The sensor data can also be downloaded remotely using the \texttt{Phyphox} URL protocol (SAVE). We parametrize \texttt{Phyphox} with a custom module that acquires simultaneously the accelerometer, gyroscope, magnetometer and GPS data at their respective maximum sampling frequencies. As described below in this section, we managed to use \texttt{Phyphox} simultaneously on a fleet of 60 smartphones to remotely run acquisitions and gather data. However, we have faced several limitations inherent to \texttt{Phyphox} design. First, \texttt{Phyphox} does not continuously save the sensor data on the smartphone internal storage, such that the recordings accumulate in a buffer memory, limiting the duration of continuous recordings to a size of about 10MB without interruption. Second, \texttt{Phyphox} requires an initial physical access to the phone to activate the distant access. Third, no time synchronization protocol with a distant host is implemented. To circumvent these three limitations, we developed an Android smartphone application called \texttt{Gobannos}~\cite{Gobannos_2025}. This application allows remote access to the smartphone sensors, continuous saving of sensor data to the smartphone memory, and synchronization of the phone's clock with a remote computer with a better accuracy.

\subsubsection{Network configuration \& remote control of the fleet}\label{sec:fleetconfig}

\begin{figure}[t!]
\includegraphics[width=\columnwidth]{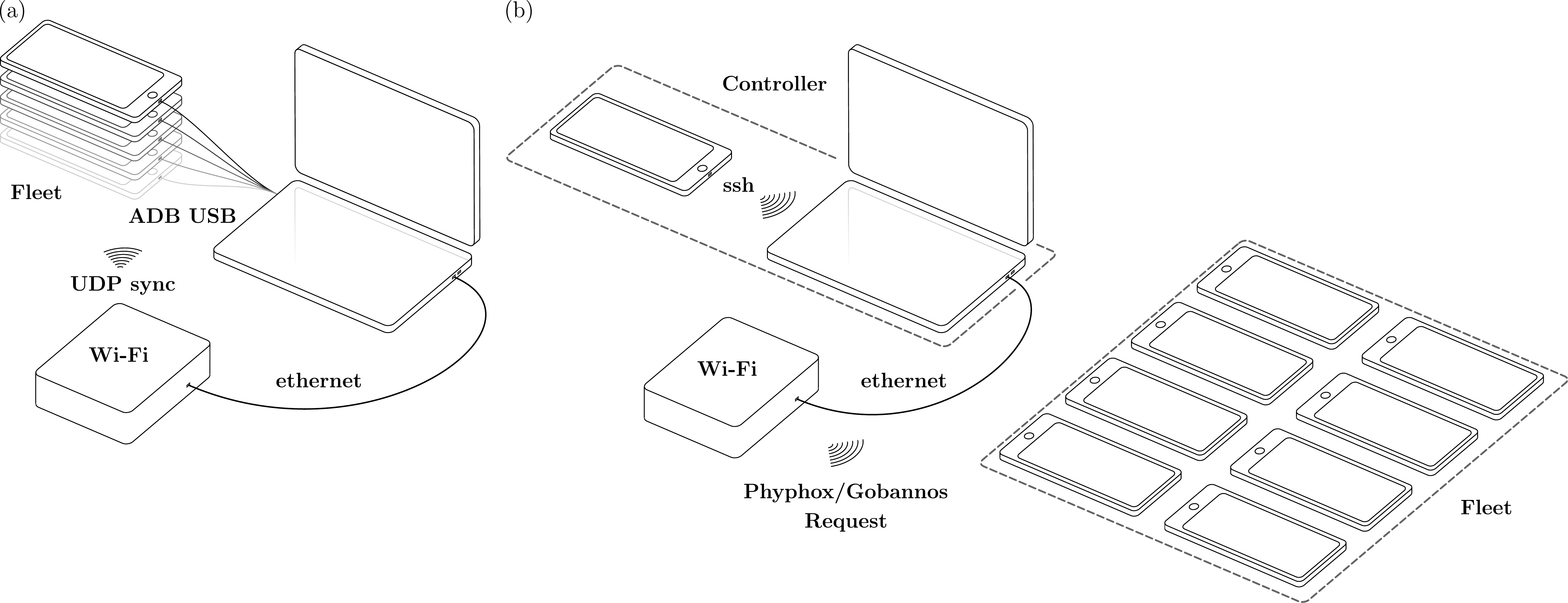}
\caption{\label{fig:Comm} Conceptual sketch for (a) fleet remote control installation protocol and (b) fleet field deployment protocol.}
\end{figure}

To operate the fleet remotely, we connect all smartphones to the same Wi-Fi network. We use a Wi-Fi broadcast system that supports at least 70 parallel connections. A laptop connected to this local network, namely the controller, is used as a DHCP server for Wi-Fi broadcast system. This controller is also used to send remote instructions to the smartphone fleet. The DHCP server attributes to each smartphone a static IP address 192.168.X.1YY, where YY is the smartphone identification number, and X is the local network identifier. The correspondence between the MAC address and the IP is saved in a text file, and used to write the configuration file of the DHCP server. The local network is disconnected from the Internet to avoid unwanted notifications from the phone OEM applications. We notice that for each smartphone, the network parameters need to be set to static IP address to avoid Wi-Fi disconnections, when connected to a local network without internet connection. Eventually, we can identify each smartphone with its static IP address. 

The scalability of the Wi-Fi broadcast system is mainly limited by network-level constraints. In practice, the maximum fleet size is set by the number of stable simultaneous Wi-Fi connections supported by the access point, as imposed by the manufacturer through hardware capabilities.

All the smartphones run on the same version of the Android operating system. In a early version, we use Android Debug Bridge (ADB) to communicate command line instructions to the smartphones from the controller. However, ADB has to be authorized on the phone, and the procedure depends on the smartphone type and the constructor. We first unlock the developer mode manually on each smartphone. To authorize ADB USB connections, Xiaomi required a Mi account associated with a personal SIM card and an email address. Each series of 3 smartphones required a different combination of SIM card and email address to be unlocked, but up to 6 smartphones could be unlocked for each email address using two different SIM cards. We then used 22 different SIM cards and 11 generic email addresses to unlock the 66 smartphones. Once connected to a Mi account, we allow the ADB USB connections, and the SIM card can then be removed. This procedure is \textit{a priori} specific to Redmi, as other constructors do not necessarily require personal information to authorize ADB USB connections. 

Prior to each day of experiment, each smartphone was connected to the controller using an USB cable, and an ADB connection was set automatically using the script \textit{adb\_usb.sh}. The script uses a second table to map the ADB USB identifier, unique to a smartphone, to the right IP address. The ADB links are stable for at least a day, as long as the smartphone is connected continuously to the Wi-Fi. The individual ADB links between the controller and the smartphones are used to start \texttt{Phyphox} remotely, check the phone state (battery level, temperature, activity), and exchange files between the controller and the smartphones. Once the \texttt{Phyphox} communication protocol is enabled, we use URL instructions to start, stop, and save the data remotely. We conclude that ADB is a versatile toolbox for controlling the smartphone state, however, it can become time-consuming. Our later implementation of \texttt{Gobannos} have made the ADB link mostly obsolete for our purpose.
    
\subsubsection{Time synchronization}\label{sec:fleetsynchro}

The time synchronization of a distributed network such as a smartphone fleet is challenging. For synchronizing clocks in a distributed network, the Network Time Protocol (NTP) was introduced and normalised in 1985 and is used to synchronize computer clocks over the Internet~\cite{mills1991internet}. For better performance, the Precision Time Protocol (PTP)~\cite{eidson2002ieee} was later introduced. Both protocols use information exchange between a server and a client, to estimate the time delay between their internal clocks. The delay introduced by the communication time between the two machines is deduced, by assuming that the communication time is symmetric, and does not depend on time. The typical precision of a NTP protocol on a local network is of the order of 1 ms, while a PTP protocol can achieve $\mu$s precision under optimal conditions. 

To test the clock synchronization using standard protocol, we implemented manually a network time protocol (NTP) and performed $10^5$ time requests between a server and 3 test smartphones using ADB time requests. Due to standard delays in Wi-Fi communication protocol, the typical duration of a time request is 100 milliseconds. Overall, the error of the NTP was found to be of few ms for a optimal Wi-Fi communication (direct line, less than 3 meter distance), but rises up to 100 ms for larger distances to the Wi-Fi access point. In optimal conditions, we used these time requests to estimate the properties of the phone clock. We found that the typical delay between two phones is of the order of 1 second, a much larger delay than the precision of the GPS clock. However, the time difference between two phone clocks is stable over time: we identify less than 10 ms of time drift between two smartphone clocks over 10 hours. A time synchronization every day was then found to be sufficient. In practice, a mechanical time synchronization gives a typical error of few ms between the smartphone clocks, and was found to be sufficient for most applications.
    
Thanks to \texttt{Gobannos}, we later implemented a time synchronization protocol based on UDP communication. The packets are sent from the controller to the phone and sent back with a typical total request time of about 2ms. The typical error on the time delay between the two clocks is  500 $\mu$s. In practice we perform 100 requests on each phone and we average the time delay over the half of the fastest time requests.\rev{We eventually achieve a typical software accuracy between two phone clocks of about 60$\mu$s}. \rev{We have checked the mechanical synchronization between the phones by using two experimental set-up described in ~\ref{sec:timecalib}. We eventually achieved a mechanical synchronization between two phones of $\Delta t = 59.5 \pm 8 \mu$s.}

\subsubsection{Error estimate and fleet reliability}

We eventually used the fleet of smartphone to test statistically the sensor accuracy. Table~\ref{table:tableSensor} summarizes the resolution, rate and average value of all sensors. The time synchronisation is common to all sensors. Note that the resolution is about ten times more accurate than the variability on the average, showing that better accuracy can be achieved by calibrating the phones. The rates present a significant variation  of about 1\% among the fleet. Eventually, our time protocol achieve a clock synchronisation of 60 $\mu$s. \\

In practical conditions and in particular in outdoor environments, the most common source of error is the reliability of the Wifi network. Starting the acquisitions indoor before deployment allows us to achieve a 100\% reliability. The ability of the \texttt{Gobannos} application to store the data locally at regular times also guarantees a 100\% reliability on data storage. Starting, stopping and retrieving data from the phone in outdoor condition introduces additional sources of error, and we practically achieve a 95\% success rate, the remaining phone requiring manual adjustment (restart \texttt{Gobannos}, Wi-Fi connection troubleshoot). The main error encountered originate from unstable Wi-Fi connection. The smartphone fleet has been used at a temperature ranging from -25 $^\circ$C to 30 $^\circ$C, and shows equal reliability in this range. Note that for negative temperature, each phone was embedded in a 3D printed box, that insulates thermally the phones, reaching typically a temperature $15^\circ$C higher than the external temperature.

\begin{table}[h]
\begin{tabular}{|c |c |c |c|} 
 \hline
 & accelerometer & gyroscope & magnetometer \\ [1ex] 
 \hline
 \hline
Resolution$^*$ &  0.0012 $\mathrm{m/s^2}$ & 0.0011 $\mathrm{rad/s}$ & 0.15 $\mathrm{\mu T}$\\ [0.5ex]
 \hline
Rate &  403.23 $\pm$ 4.35 Hz & 403.23 $\pm$ 4.35 Hz & 50.0 $\pm$ 1.31 E-5  Hz \\ [0.5ex]
 \hline
Average & 9.82 $\pm$ 0.02 $\mathrm{m/s^2}$ & 0.02 $\pm$ 0.02 $\mathrm{rad/s}$ & 104.76 $\pm$ 72.32  $\mathrm{\mu T}$ \\ [0.5ex]
 \hline
Time sync & \multicolumn{3}{l|}{UDP protocol: 60 $\mathrm{\mu s}$. Mechanical protocol: 59.5 $\pm$ 8 $\mathrm{\mu s}$} \\ [0.5ex]
 \hline
\end{tabular}
\caption{Characteristics of embedded IMU sensors for Redmi 10 smartphone model 22011119UY. Values are obtained from static test of 50 smartphones. *: data obtained from Phyphox database~https://phyphox.org/sensordb/}
\label{table:tableSensor}
\end{table}

\subsubsection{PhoneFleet application}\label{sec:pyphone}

To facilitate the remote control of the smartphone fleet, we developed a Python application called \texttt{PhoneFleet} available on github.com~\cite{Phonefleet_2024}, which implements the functionnalities described above. It includes in particular the DHCP server configuration, the creation of ADB links, the start of \texttt{Phyphox} remotely, the control of \texttt{Phyphox} or \texttt{Gobannos} acquisitions through the URL protocol (run, stop clear and save). To optimize time delays in the execution of a command on multiple phones, \texttt{PhoneFleet} uses asynchronous Python functions, as well as multi-threading to run tasks in parallel, on subparts of the smartphone fleet. \texttt{PhoneFleet} currently uses a graphical interface based on PyQt5 with buttons and tabs to organize the functions in themes, and run the commands on the selected phones. The application format may evolve in the near future to improve portability and maintenance capacity. 

\section{Applications}\label{sec:applications}

Here we present two applications of the smartphone fleet as time synchronized IMU sensors. The first section is devoted to the oscillation of a smartphone chain immersed in a turbulent flow. Each phone is then used both as the object of interest and the measuring device. The second section is devoted to the use of smartphones as local wave buoys, placed on floating sea ice. By recording the acceleration, angular velocity and earth magnetic field, we manage to measure the wave amplitude at different locations.

\subsection{Pendulum chain in turbulence}\label{sec:chainPendule}

The environmental artist Ned Kahn's artworks \textit{Kinetic Facade} are building facades covered by thousands of aluminium pendulum plates that oscillate harmoniously in the wind, creating regular patterns of ripples that resemble a fluttering flag~\cite{shelley2011flapping} or sea waves~\cite{drazin2002introduction,perrard2019turbulent}. To gain in-depth understanding of such phenomena, recent experimental investigations by Zhang and Perrard~\cite{zhang2024large} were conducted on oscillation measurement of a 1 meter chain of pendulum plates confronting a turbulent flow by camera imaging. They have evidenced wave-like advective patterns of pendulums oscillations, similar to Ned Kahn's artworks.
By spectral analyses in spatiotemporal Fourier space of the pendulum oscillations, it has been shown that these moving patterns could emerge in a turbulent flow as a result of two distinct mechanisms. One can be described as a resonant response of each pendulum near the pendulum's natural frequency of oscillation, and the other one attributed to the direct response of the pendulums to the turbulent fluctuations. The maximum response is reached at the intersection between the two dispersion relations.


To upscale the observable pendulum plate dynamics to a larger flow dimension comparable to that of a real facade, we used smartphones as rigid plate pendulums to measure spatiotemporal pressure fluctuations. We constructed a ten-meter-long chain of 60 uniformly spaced smartphones, each one hinged to a common rod along its top edge, allowing free oscillation around the $x$-axis. The built-in sensors, including accelerometers and gyroscopes, enable the determination of the instantaneous acceleration of each smartphone. This setup provides highly temporally resolved measurements of oscillations and is space-efficient, especially where imaging techniques using cameras reach their limits.

\begin{figure}[t!]
  \includegraphics[width=0.95\columnwidth]{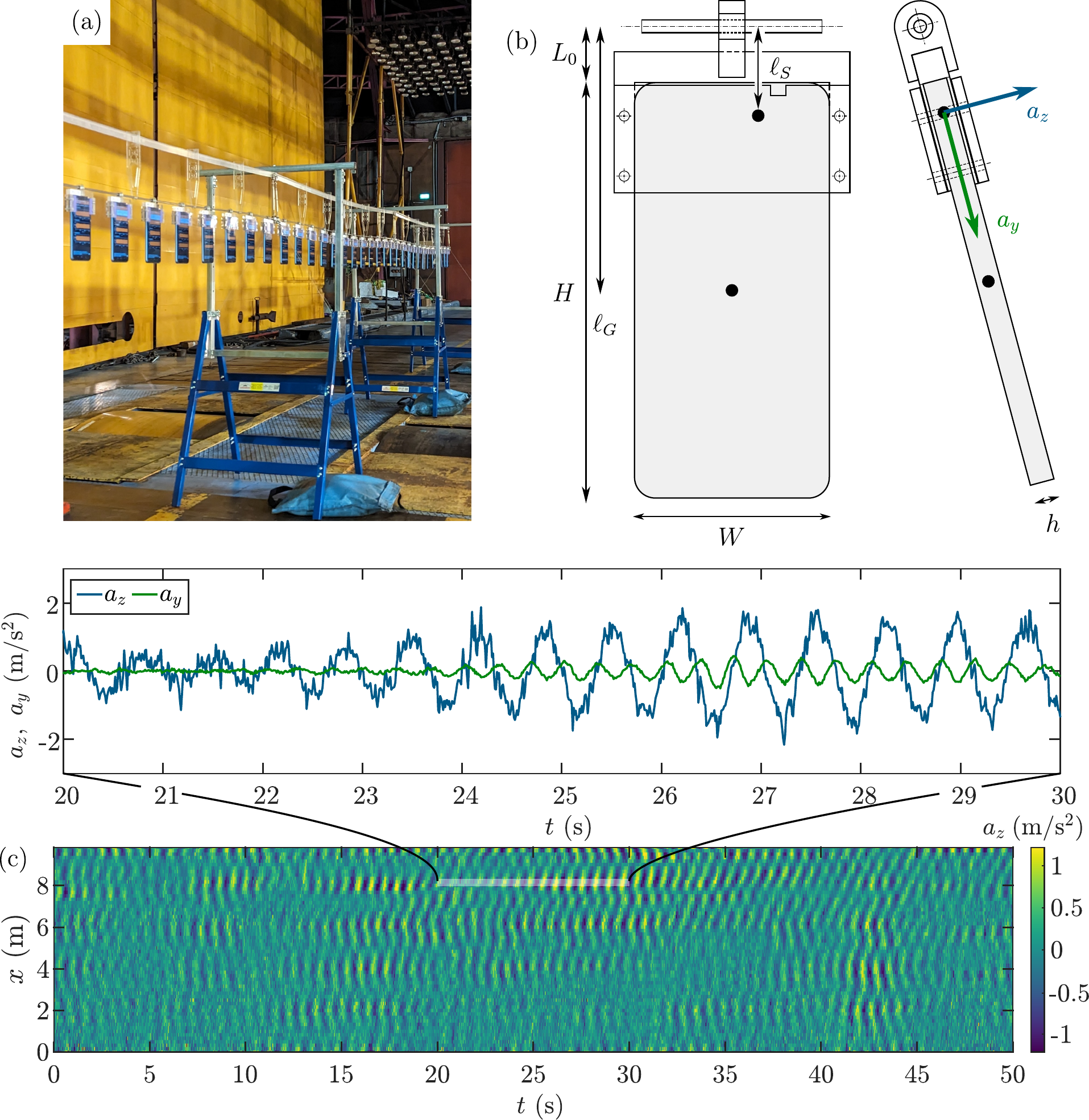}
  \caption{\label{fig:Soufflerie} a)  Image of the wind tunnel measurement setup. The chain of smartphones is placed in the symmetry plane of the measuring section, in suspension with metal trestles. b) Technical drawing of the rod-mount suspension system and its assembly. $\ell_S=50.6$ mm and $\ell_G=111.8$ mm indicate the rod to the accelerometer distance and the rod to the center of gravity distance. c) Spatiotemporal chart of the tangential acceleration $a_z$ of the chain of telephones, for a wind speed $U=9.3$ m/s. Inset: temporal signal of the tangential $a_z$ (blue) and radial (green) acceleration $a_y$ of the smartphone $\#50$ located at around $x=8$ m. Both the signal duration and smartphone position are indicated by the white dashed line.}
\end{figure}

Measurements of the chain oscillation in the wind flow were conducted in the large low speed S6 wind tunnel at Institut A\'erotechnique Saint-Cyr l'\'Ecole~\footnote{https://iat-en.cnam.fr/}. The chain of smartphones was placed in the symmetry line of the facility's test section, as shown in Fig.~\ref{fig:Soufflerie}(a). The free-stream wind speed $U$ is varied in the range 2-10 m/s, allowing an interaction of a fully established turbulent flow with the smartphone chain. The instantaneous and time-averaged statistics of the wind flow speed were measured independently via calibrated hotwire probes and dynamic pressure sensors.

In the following, we aim to characterize the spatio-temporal dynamics of the wavy patterns along the smartphone chain, based on accelerometer data. Fig.~\ref{fig:Soufflerie}b shows the assembly of one smartphone as a pendulum oscillating around the $x$ axis. The accelerometer is located at a distance $\ell_S = 50.6$ mm from the rod center axis. Assuming that the smartphone oscillates only around the $x$ axis, the angular motion of a single smartphone in a non Galilean frame of reference induces a tangential and a radial accelerations $a_z$ and $a_y$ given by~\cite{dauphin2018physical}:

\begin{eqnarray}
a_z &=& -g\sin\theta - \ell_S\ddot{\theta}, 
    \label{eq:vecA}\\[6pt]
a_y &=& g\cos\theta + \ell_S\dot{\theta}^2,
    \label{eq:vecB}
\end{eqnarray}

The tangential component $a_z$ corresponds to the acceleration measured by the sensor along the local $z$ axis. For a simple pendulum, both contributions in Eq.\ref{eq:vecA} oscillate at the natural frequency of the pendulum $\omega_0=9.37$ rad/s, which we determined empirically from free oscillation measurements. However, the radial component $a_y$ is dominated by a second-harmonic response at $2\omega_0$ (arising from the $\cos\theta$ and $\dot{\theta}^2$ terms in Eq.~\ref{eq:vecB}). Although $a_z$ and $a_y$ are nonlinear functions of the oscillating angle $\theta(t)$, their form depends only on the local motion and not on the global spatial pattern of the smartphone chain. Consequently, in the following we characterize the spatiotemporal chain dynamics directly from the raw accelerometer signals.


Fig.~\ref{fig:Soufflerie}c shows the spatiotemporal chart of the tangential accelerations $a_z$ measured by the chain of smartphones. From left to right of the chart, we observe wavy patterns propagating across the chain with an approximately constant phase speed. We also observe interference patterns with reflected waves that propagate upstream. 

\begin{figure}[t!]
\begin{center}
\includegraphics[width=.7\columnwidth]{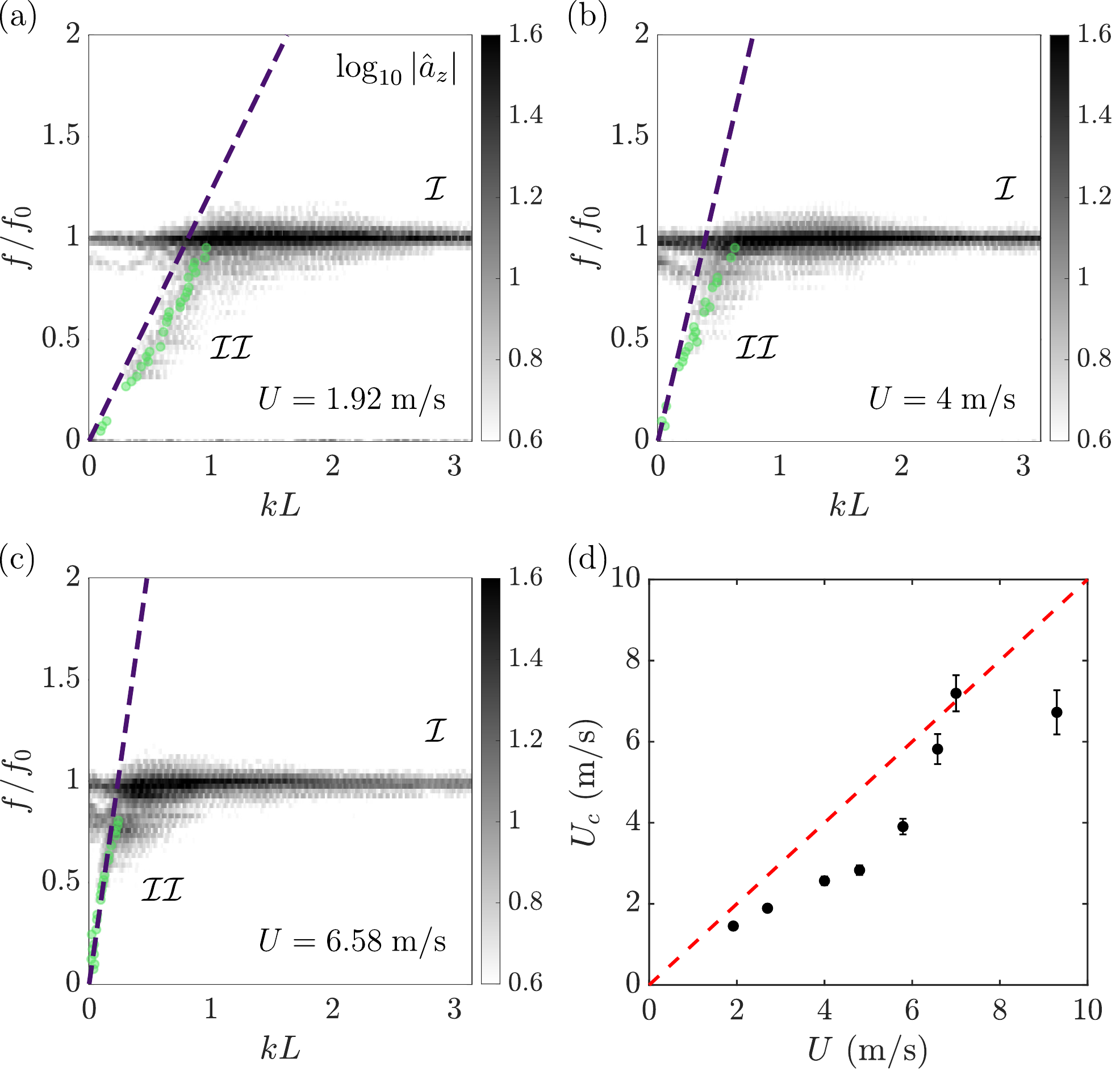}
  \caption{\label{fig:RD_soufflerie} Spatiotemporal spectrum of the normalized tangential acceleration $\log_{10} | \hat{a}_z | (f/f_0, kL)$ of the telephone chain for three increasing wind speeds (a): $U=1.92$~m/s, (b): $U=4$~m/s, (c): $U=6.58$~m/s. Frequency and wavenumber normalized by the natural frequency of oscillation $f_0$ and the smartphone spacing distance $L$. Local maxima of $|\hat{a}_z|$ on the dispersion relation $II$ for the low frequency range $f<f_0$ are indicated by green circles. Purple dashed lines correspond to the slopes given by the mean wind speed $U=2\pi f/k$. (d) Convection velocity $U_c$ measured by linear slope fit on the dispersion relation $II$ as a function of the wind speed. }
 \end{center}
 \end{figure}

We now characterize the wave dynamics of the smartphone chain in the frequency and wavenumber domain. We perform the two-dimensional discrete Fourier transform in space and in time to convert the tangential acceleration $a_z(x,t)$ in physical space into $|\hat{a}_z(f,k)|$ in spectral space. For each discrete frequency $f_i$, the spectral amplitude $|\hat{a}_z(f_i,k)|$ is normalized by its maximum over all wavenumbers, ensuring values between 0 and 1. Fig.~\ref{fig:RD_soufflerie}(a-c) shows the frequency-wavenumber spectrum of the magnitude of the tangential acceleration $| \hat{a}_z |$ for three increasing wind speeds. We observe two distinct dispersion relations governing the oscillation dynamics, consistent with our earlier observations on Ned Kahn’s \textit{Kinetic Facade} artworks~\cite{zhang2024large}. Branch \textit{I} corresponds to oscillations of the smartphone chain around the natural frequency $f/f_0=1$, marginally dependent on wavenumber $k$. We attribute this dispersion relation to a resonant response mechanism of a single smartphone to the wind's turbulent fluctuations in the vicinity of the natural frequency $f_0$. Branch \textit{II} follows a linear dispersion relation through the origin ($f/f_0=0$, $kL=0$) and describes the telephone oscillations at low frequencies $f < f_0$. These motions can be understood as a passive response to coherent turbulent structures convected by the wind at a characteristic velocity $U_c$, given by the slope of the branch. For comparison, the wind free-stream velocity $U$ is plotted as dashed purple lines $\omega=U k$, showing that the slope of Branch \textit{II} increases with wind speed but remains systematically smaller than $U$.

In order to quantify $U_c$, we subsequently extract the local maximum of $|\hat{a}_z(f,k)|$ over wavenumber $k_c$ (green circles) for a given frequency, and perform a linear fit along Branch \textit{II} by $\omega = U_c k_c$. The resulting convection velocity as a function of wind speed is shown in Fig.~\ref{fig:RD_soufflerie}(d). We see a clear positive correlation between the wind velocity and the velocity $U_c$ characterizing the advection of the turbulent pressure fluctuations along the chain. On average, the ratio $U_c/U=0.75\pm 0.04$, consistent with 0.8 in our previous study on the reduced chain of pendulums~\cite{zhang2024large}, and comparable to the convection velocity of wall-pressure fluctuations in a zero-pressure-gradient turbulent boundary layer~\cite{choi1990space}. We notice that the ratio $U_c/U$ approaches or exceeds 1 for higher wind speeds ($U=6.58$ m/s and 7 m/s), possibly due to the larger scatter of the local maxima in this regime. To summarize, by exploiting the embedded accelerometers of the smartphone chain, we have measured the spatiotemporal Fourier spectrum of the pendulum oscillations and revealed the emergence of a two-branch dispersion relation under wind forcing. Branch \textit{I} arises from the resonance of individual smartphones, each responding locally to turbulent pressure fluctuations near their natural frequency $\omega_0$. Branch \textit{II}, on the other hand, originates from the collective motion of the chain, where the pendulums respond coherently to the advection of large-scale turbulent structures. These underlying mechanisms explain the coexistence of a nearly flat resonant branch and a linear convective branch in the spectrum. From the latter, we reliably extracted the convection velocity of large-scale turbulent pressure fluctuations. The measured convection velocity scales as a constant fraction of the free-stream velocity, consistent with observations in turbulent boundary layers. This validates our approach as a robust method for large scale measurements of flow-induced oscillations of structures, at dimensions directly comparable to Ned Kahn's \textit{Kinetic Facade} artworks.

\subsection{Wave buoys}\label{sec:buoys}

 \begin{figure}[t!]
 \includegraphics[width=0.95\columnwidth]{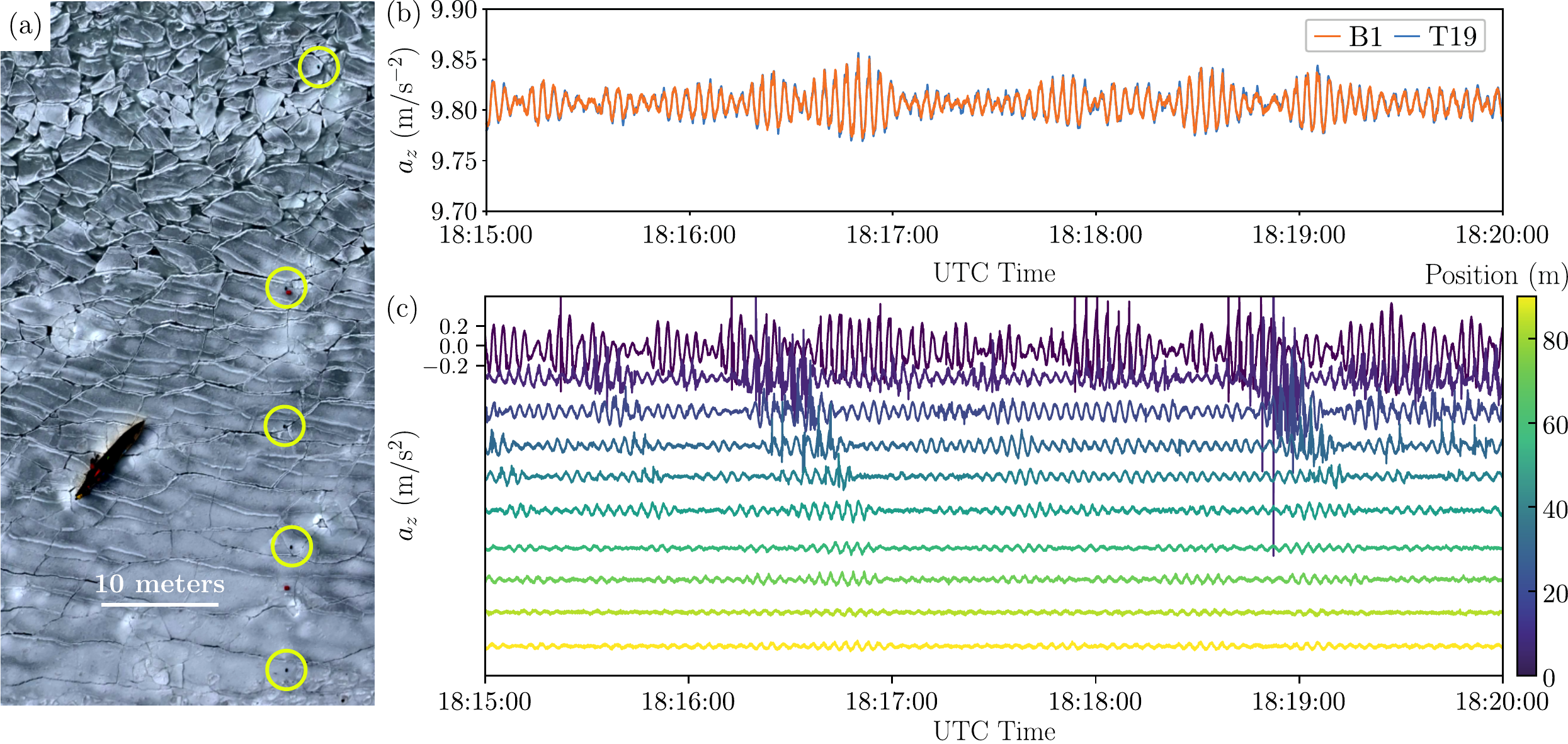}
 \caption{
    \label{fig:spatiotemp}
    Test case \#2: wave propagation under ice. 
    (a) Top-view photography of the experimental situation. The ice edge is at
    the top, and the ice becomes more continuous when moving away from it.
    Orange boxes (as seen next to the second phone, and between the fourth and
    fifth phones, counted from the top of the image) are wave buoys. Recording
    smartphones are highlighted with circles.
    (b) Comparison between vertical accelerations as measured by a phone and a
    wave buoy. The mean of the buoy signal as been adjusted to that of the
    phone to better visualize the agreement; the instrument bias is
    0.035~m/s$^2$.
    (c) Temporal signals of ten smartphones placed 10~m apart, as described in
    the text. The distance is measured from the phone closest to the ice edge.}
 \end{figure}
 
The smartphone fleet can also be used as a network of autonomous IMU systems in outdoor conditions. We perform a proof of concept during a field campaign on wave-sea ice interaction in Rimouski, Qu\'ebec, Canada, in February 2024, where the smartphones were deployed as local wave buoys to record the motion of ice induced by gravity waves \cite{kuchly2025integrated}. Smartphones can aptly be used to record the motion of the ice induced by gravity waves, as an array of smartphones can provide spatial and temporal information on wave propagation and attenuation. Custom cases were 3D-printed to shield phones from the environmental condition, ensuring their temperature remains within working range.

On 23 February, we noticed wave activity (thickness of 16~cm) in Mercier Cove, in the estuary of the Saint Lawrence river. We placed ten phones close to the forming ice edge, on a straight line, aligned with the direction of wave propagation, regularly spaced 10 meters apart. We recorded GPS coordinates, as well as acceleration, gyroscope and magnetometer signals. The recordings overlapped in time for about thirty minutes, and we only present the exploitation of vertical accelerations in this section, as a proxy for wave height. We show a top-view of the cropped
situation, photographed by a drone, in Fig.~\ref{fig:spatiotemp}(a).

The phone-recorded positions proved to be accurate, within about 5~m, to positions recorded with a multi band hand-held GPS device. Unfortunately, due to hardware limitation, the recorded positions do not vary in time from the moment the phones were laid on the ice, which prevents us from extracting more precise positions through averaging. The GPS time sampling is 1Hz, and the minimum measured displacement is 12 cm.

Two of these phones were doubled with wave buoys developed by P. Sutherland from Ifremer~\cite{Sutherland_2018}, themselves equipped with geolocation and accelerometers, similar to the model described by Guimar\~aes \textit{et al.}~\cite{Guimaraes_2018}. After the time-synchronization of the phones signals, we set their time origin by finding the lag maximizing the cross-correlation between the acceleration signals of the phone (\#T19) and the buoy (\#B1) situated at the same location. We \rev{show an excellent agreement between the wave buoy signal and the smartphone accelerometer sensor (Fig.~\ref{fig:spatiotemp}(b)).} In Fig.~\ref{fig:spatiotemp}(c), we show the time evolution of the vertical acceleration as recorded by 10 phones at increasing distance from the ice edge. Qualitatively, we confirm the synchronization of the signals by observing individual wave packets propagating in space. These recordings reveal the attenuation of the wave amplitude as the wave propagates under partially fragmented ice.

\begin{figure}[t!]
     \includegraphics[width=0.95\columnwidth]{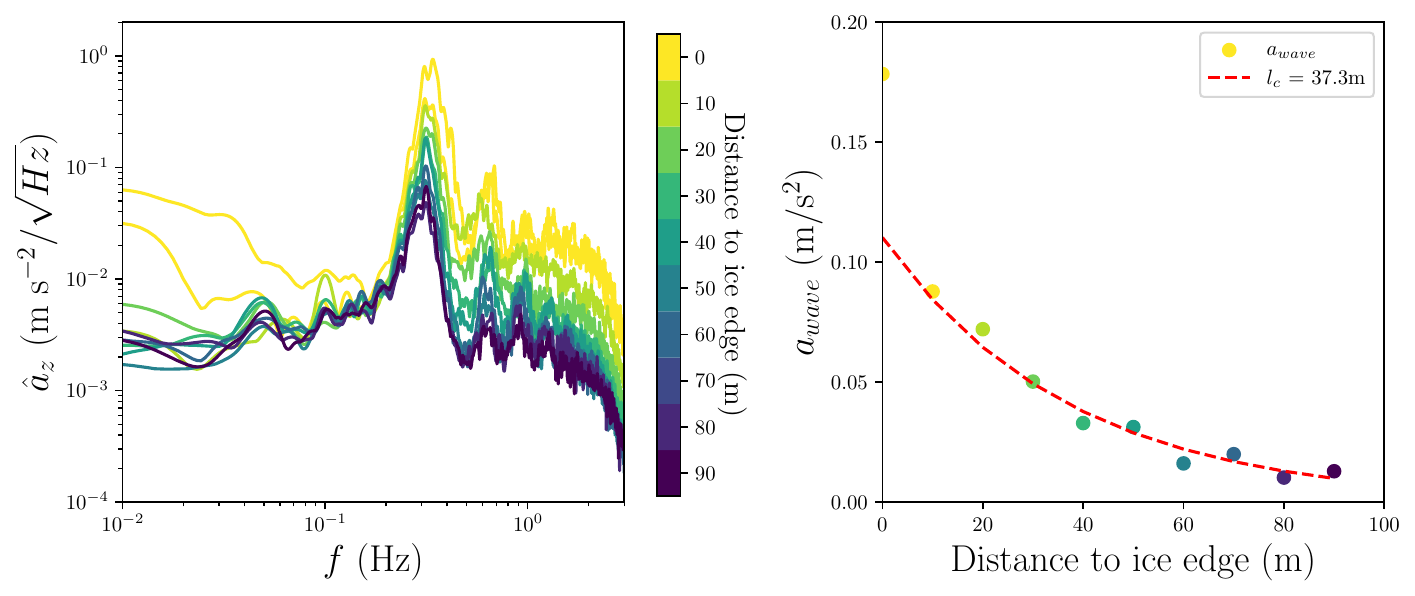}
     \includegraphics[width=0.95\columnwidth]{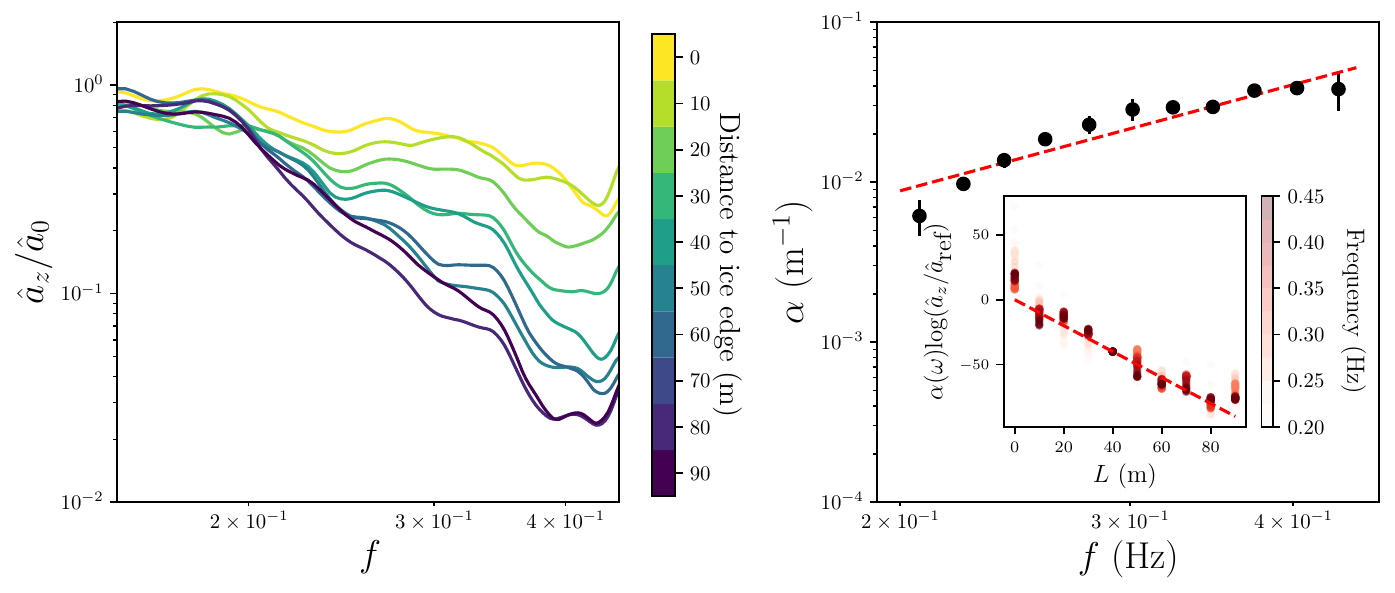}
     \caption{Measured in-ice wave propagation by phone accelerometers placed at increasing distances from the ice edge, in the direction of wave propagation.
        (a)~Temporal spectra $\hat a_z$ of the vertical acceleration. The color encodes the distance to the ice edge. The peak at $f_0 = 0.3 \pm 0.02$Hz corresponds to the most energetical waves.
        (b)~Acceleration amplitude associated with wave-induced vertical motion inferred from the acceleration spectra, as a function of the distance from the ice edge (points), and exponential fit
        (dashed line) with an attenuation length $\ell_c = 37.3$~m.
        (c)~Zoom in view of the spectra normalised by the spectrum of the first phone $\hat a/\hat a_0$, showing the fast decay of the wave energy at higher frequency.
        (d)~Measure of the attenuation coefficient $\alpha(f)$ per frequency, in the range $f \in [0.2,0.45]$Hz, extracted from an exponential fit of $\hat a/\hat a_0$ as a function of $L$. We found approximately $\alpha = \alpha_0 f^{-2.2}$ with $\alpha_0=0.31$/m from a fit by a power law (red dashed line). Insert : Normalised acceleration amplitude to check the validity of the exponential decay model in the entire frequency range. 
    }\label{fig:seaice}
\end{figure}

\rev{Fig.~\ref{fig:seaice} shows the temporal Fourier spectrum of $a_z$ for the 10 smartphones, where the color corresponds to the distance to the ice edge. All the spectra present a maximum at $f_0 = 0.3 \pm 0.02$Hz, and significant wave energy in the principal peak between 0.2 to 0.45~Hz, corresponding to the coastal swell propagating under the ice. Note that higher frequencies component of the spectrum are a sum of non linear contribution of the main peak, and energy at higher frequency. From these spectra, we compute a proxy of the wave amplitude $a_w$ as the standard deviation of $a_z$, filtered in the frequency range [0.1,5]~Hz. The wave amplitude as a function of the ice edge is shown in Fig.~\ref{fig:seaice}b and is well fitted by an exponential decay, $a_w = a_0 \textrm{exp}(-x/\ell_c)$, where $x$ is the distance to the ice edge, and $\ell_c = 37.3$m is a characteristic decay length. The wave amplitude $A$ can be inferred from the acceleration signal as $A ~\sim 2 (\pi f_0)^2 $, corresponding to $A$ ranging from 4cm at the ice edge, to 3mm at 90m from the ice edge. It is remarkable to measure the wave amplitude and the decay for such small wave amplitudes. We conducted additional analysis to extract a damping coefficient as a function of the wave frequency. Figure~\ref{fig:seaice}c) shows a zoom in view of the spectrum, normalised by the wave spectrum of the first phone located at the ice edge. From these spectrum, we fit an exponential decay $\hat a$ = $\hat a_0 exp(-\alpha(f) L)$, and we extract the damping coefficient $\alpha$, shown in figure~\ref{fig:seaice}d as a function of the wave frequency. We checked that the exponential fit was a valid model, by representing $\alpha(f) \textrm{log}(\hat a_z/\hat a_{ref})$ as a function of the distance $L$ to the ice edge. The result is shown in the insert of figure~\ref{fig:seaice}d, where the color encodes the wave frequency. We found a good agreement with an exponential decay, all the curves lying in the vicinity of a linear segment of slope -1 (red dashed line). Unfortunately, the ice parameters could not be inferred from these measurements, as the range of excited frequency is too small to be sensitive to the ice properties (thickness, Young's modulus, Poisson ratio).}

\rev{These measurements show that smartphones can be deployed and used reliably to access wave data in outdoor conditions, and characterized typical properties of ocean waves, such as wave spectrum, or spatial decay. Although we did not manage to extract ice mechanical properties from these preliminary experiments, we believe that further works focusing on a higher frequency range combined with the use active sources, could provide an estimate of the Young's modulus or the ice thickness from wave propagation.}

\section{Discussion}

We have presented a methodology that transforms homogeneous smartphones into a calibrated, time-synchronised, and remotely coordinated large-scale sensor fleet for physical field measurements. By achieving microsecond-level synchronization, enabling centralised control and parallel data acquisition across tens of devices, this approach supports coherent multi-point measurements of rapidly evolving physical phenomena such as mechanical waves by wind turbulence or wave–ice interactions. Compared with established wireless sensor networks (WSNs) and IoT-based frameworks, our system overcomes several long-standing limitations: typical WSN synchronization protocols are constrained by heterogeneous devices and hardware clock-drift accumulation, which generally limit precision to the millisecond or tens-of-microseconds range. In contrast, our smartphone fleet benefits from homogeneous, well-characterised sensors, stable clocks, and a dedicated synchronization protocol that consistently achieves µs-level alignment, enabling coherent spatiotemporal reconstruction of dynamic fields. This positions our framework between ad-hoc smartphone sensing and specialised professional distributed instrumentation, offering the precision and cost-effectiveness required for dense field deployments. Overall, our results demonstrate that widely available smartphones can form a reliable scientific instrument for distributed measurements, opening new possibilities for large-scale, accessible field experiments in environmental, mechanical, and geophysical systems.

\section{Conclusion}\label{sec:conclusion}

We show that a fleet of smartphone can be controlled from a single computer, to perform spatial measurements of mechanical vibrations. We first evaluate the sensors sensitivity and accuracy for a single smartphone and precisely determine the sensor locations. Based on existing protocols for single smartphones, we introduced a new architecture based on two softwares, \texttt{Gobannos} and \texttt{Phonefleet}. \texttt{Gobannos} is installed on each smartphone, and allows to remote-control the entire acquisition process using url requests. \texttt{Phonefleet} is deployed on the controller, to send the instructions to the fleet, and gather the results. The combination of the two, with an appropriate local Wi-Fi network opens new avenue for smartphone-based multi-physics instrumentation. We eventually demonstrate the capabilities, by conducting experiments with the smartphone fleet in two model situations. The first experiment focuses on the large-scale physical measurements of turbulent fluctuations using smartphones as pendulums. The second experiment focuses on wave-ice floe interactions using smartphones as local wave buoys. Altogether, this work paves the way for low cost and eco-friendly solutions for large-scale scientific studies. \\

\section*{CRediT authorship contribution statement}
Jishen Zhang: Conceptualization, Methodology, Investigation, Formal analysis, Writing - Original Draft. Nicolas Mokus: Investigation, Writing - Review \& Editing. Jules Casoli: Software, Writing - Review \& Editing. Antonin Eddi: Conceptualization, Methodology, Software, Investigation, Formal analysis, Writing - Original Draft. Stéphane Perrard: Conceptualization, Methodology, Software, Investigation, Formal analysis, Writing - Original Draft.\\

\textit{acknowledgments} We thank B. Auvity, S. Kuchly, B. Lafoux and G. Vandenhove for scientific discussions \& technical developments, as well as PMMH members C. Aracheloff, R. Baillou, A. Bouvier, S. Gom\'e, N. He,  S. Kumar Saroj, A. Mongruel,  B. Musci, P. Petitjeans, W. Reino,  A. Rivi\`ere,  J. Tampier, V. Thievenaz, L. Tuckermann and D. Vandembroucq for providing their SIM card to unlock each 3 smartphones of the fleet. We thank C. Robert and the technical staff of IAT Saint-Cyr-l'\'Ecole for their help with the wind tunnel experiment. We also thank T-CAPES Initiative for scientific and financial support. This work has benefited from the financial support of Mairie de Paris through Emergence(s) grant 2021-DAE-100 245973, of the Agence Nationale de la Recherche through grants MSIM ANR-23-CE01-0020-02 \& LASCATURB ANR-23-CE30-0043-03, and of the PSL University through a Junior Fellowship 2022-305.

\appendix{}
\section{Sensor calibration and accuracy tests}\label{sec:gyrocalib}
\subsection{Gyroscope}

\begin{figure}[htbp]
 \includegraphics[width=\columnwidth]{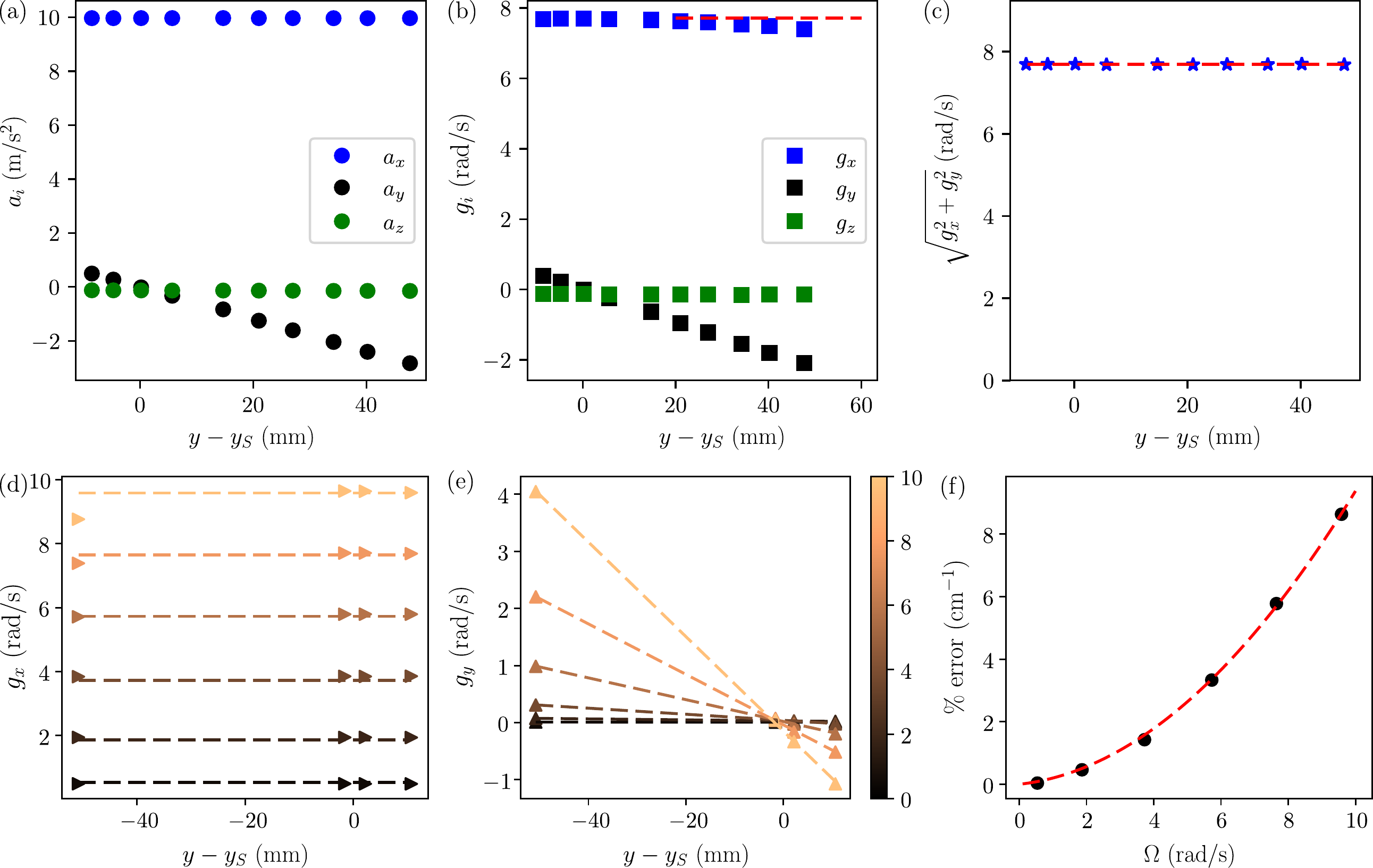}
  \caption{\label{fig:calib_g_a} Calibration of accelerometer and gyroscope. The measurement depends on the axis of rotation. When the smartphone rotate around the position of the gyroscope, it actually measures the angular velocity within 1\% of error. When the smartphone rotates around an off-centered axis, the measured values depend on the distance $r$ to the axis of rotation. This discrepancy originates from hardware limitations. (a) 3-axis acceleration signals recorded for a rotating phone. (b) 3-axis gyroscope signals recorder in the same rotating configuration. (c) Total gyroscopic signal $g = \sqrt{g_x^2+g_y^2}$ in this rotating configuration.
  (d) Measured angular velocity along $y$ from $g_x$ signal, as a function of the position of the smartphone with respect to the rotation axis, compared to the expected value. The imposed angular speed $\Omega$ is color-coded. e) Measurement along $g_y$, corresponding to a rotation around $x$ that does not physically occur. This corresponds to a systematic error, which increases linearly with $r$, and increases with rotation speed. From these measurements, we infer that the minimal error is reached when the smartphone rotates around the position of the gyroscope, we find $r_g = 31.3 \pm 1$mm, different from the accelerometer location. f) Percentage of error $g_y/(\Omega r)$ as a function of the angular velocity $\Omega$, in cm$^{-1}$.}
 \end{figure}
 
\rev{The gyroscope accuracy has been tested using the rotating table introduced in section~\ref{sec:MEMSlcs}, the accelerometer data serving as a reference. The phone rotates along the $x$ axis at an angular frequency $\Omega = 7.7$~rad/s, varying the position of the phone with respect to the rotation axis (sketch of Fig.~\ref{fig:acceleroCentre}(b)). Fig.~\ref{fig:calib_g_a}(a) shows the average acceleration as a function of the relative sensor position $y-y_S$ to the rotation axis. The three colors corresponds to the components $a_x$, $a_y$ and $a_z$ of the acceleration. We recover that the sensor measures the acceleration of gravity along the $y$ axis for all sliding positions, the centrifugal force along $x$, and the $z$ component is identically zero.
 Fig.~\ref{fig:calib_g_a}(b) shows the average angular velocity components $g_x,g_y$ and $g_z$ measured simultaneously by the gyroscope, as a function of the relative smartphone position $y-y_S$. The main rotation is indeed found along $y$, and the value agrees quantitatively with less than 0.5\% of error when the accelerometer sensor is on the rotation axis ($y = y_s$). However, we also measures a linear variation of $g_x$ as a function of $y-y_S$ and a small (quadratic) decrease of $g_y$. These measurements do not correspond to a true phone rotation as evidenced by the accelerometer signals, but rather to a limitation of the gyroscope sensor. In this specific case of a known rotation axis, the systematic error can be compensated, by noticing that $g = \sqrt{g_x^2+g_y^2}$ is conserved for all tested smartphone positions $y-y_S$ (Fig.~\ref{fig:calib_g_a}(c)), and correspond to the imposed angular frequency $\Omega$. The same behaviour is observed for all three directions of rotation. This discrepancy originates from the limitation of the measurement technique implemented in the gyroscope sensor. In practice, the gyroscope is a vibrating structure gyroscope (VSG) which is based on the Coriolis effect applied on a vibrating mass. As a consequence, in a non Galilean frame of reference, the gyroscope measures a combination of inertial acceleration and Coriolis force, and the sensor reading deviates from the true angular frequency, both in magnitude and in direction. 
 To further test the gyroscope sensor limitations, we perform additional measurements with increasing rotation rate, at four different phone locations $y-y_S$ for a rotation around the $x$ axis of the phone. Fig.~\ref{fig:calib_g_a}(d) shows the main angular velocity component $g_x$ as a function of $y-y_S$, for different rotation rates (color coded). These measurements confirm that for small $\Omega$ (typically $\Omega<5$ rad/s), we recover $g_x \sim \Omega$. For larger rotation rate, we observe a significant discrepancy, which depends on the smartphone location. In particular for $|y-y_S|$ = 50 mm the error grows to 10\% at $\Omega = 10$~rad/s. The systematic error is best evidenced by the signals of $g_y$ (Fig.~\ref{fig:calib_g_a}(e)), which should be identically zero. For all rotation speeds, the systematic error is linear with the distance $r$ between the sensor and the axis of rotation, and the slope increases with $\Omega$. From these measurements, we infer that the gyroscope sensor is located at the position $y_g = 31.3 \pm 1$~mm along the $y$ axis, very close to the accelerometer sensor. Eventually, we extract the slope $\alpha_g$ of the error $g_y = \alpha_g (r-r_g)$, and we compute a relative error, $\Delta g = \alpha_g/\Omega$, which is measured in $\%/cm$, as it is an error proportional to $r$. The error is represented in Fig.~\ref{fig:calib_g_a}(f) as a function of $\Omega$. For a period of 1s corresponding to $\Omega = 2 \pi$ rad/s, we found 4\% of error per centimeter. Among the sources of error, we identify (i) mechanical coupling between the $x$ and $y$ gyroscopes, which plays a symmetrical roles. It would explain why the orientation of the rotation vector is modified but the total angular velocity $\Omega$ is conserved at moderate rotation and moderate distance to the axis of rotation. The assumption of pure Coriolis force measured by $g_i$ fails for increasing velocity and acceleration of the gyroscope sensors, it adds at least two other sources of error : (ii) The sensor also measure inertial forces, such as centrifugal force, which is proportional to $r$. (iii) The assumption of a purely oscillatory motion imposed to the sensor by the ship is no longer valid, the velocity of the gyro with respect to the Galilean frame of reference is no longer negligible. For a rotation at constant speed, all these sources of error are proportional to the distance $r$ to the axis, but depend differently on the angular velocity $\Omega$. For oscillatory motions and a distance $r$ to the center of rotation varying in time, these systematic sources of errors would be difficult to disentangled. We conclude that in practice, the gyroscope of the Redmi10A is reliable only to measure pure rotational motions around the gyroscope sensor, and cannot be used in general in combination with the accelerometer to decompose arbitrary superposition of translational and rotational motions. However, in some limit cases, in particular when the center of rotation is known, we expect the gyroscope to be reliable. One may refer to Fig.~\ref{fig:calib_g_a}(f) to evaluate the systematic error as a function of both the angular frequency and the distance $r$ between the sensor and the axis of rotation.}
 
\begin{figure}[t!]
\begin{center}
 \includegraphics[width=0.5\columnwidth]{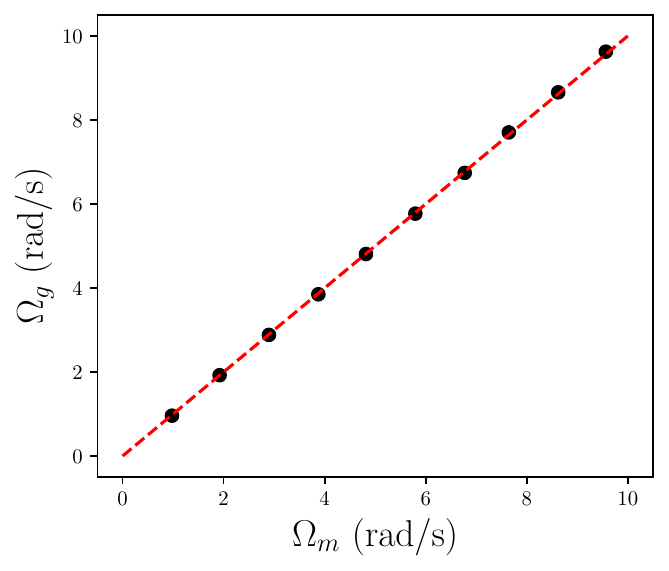}
  \caption{\label{fig:calib_g_FP} Test of the gyroscope accuracy on a Fairphone 4. Averaged angular velocity $\Omega_g$ measured with the gyroscope as a function of the angular velocity $\Omega_m$ measured with the magnetometer using $\Omega_m = 2 \pi f_0$, where $f_0$ is the frequency of oscillation of the horizontal components of the magnetometer. The Red dashed line corresponds to a line of slope 1, showing a quantitative agreement between the two measurement techniques.}
  \end{center}
 \end{figure}

\rev{For comparison, we test another smartphone, the Fairphone 4 (FP4), which also embeds an accelerometer, a gyroscope, and a magnetometer. We conduct the same series of tests on the FP4. The smartphone is mounted on a rotating table, as sketched on figure~\ref{fig:Sketch}b), rotating at a constant angular velocity $\Omega$.
 We first measure the angular velocity $\Omega_m$ with the magnetometer using the frequency of oscillation $f_0$ of any component of the magnetometer sensor. The frequency $f_0$ is extracted from the Fourier spectrum of the component $x$ of the magnetometer, and we have $\Omega_m = 2 \pi f_0$. The angular velocity is also measured from the gyroscope sensor. We measure the component of the gyroscope along the direction of the rotation axis, and we extract its temporal average $\Omega_g$, for different rotation speed. Figure~\ref{fig:calib_g_FP} shows the angular velocity $\Omega_g$ measured with the gyroscrope sensor, as a function of the angular velocity $\Omega_m$ measured with the magnetometer. The red dashed line corresponds to a line of slope 1, and it fits perfectly the data points. Therefore, the gyroscope sensor of the FP4 measures the angular velocity without any significant error up to the maximum tested rotation speed ($\Omega \sim 10$ rad/s), independent of the distance between the sensor and the axis of rotation. We conclude that the limitation of the sensors and their accuracy can vary drastically from one smartphone to another, and a benchmark of the sensors, following for instance the protocols presented in this article, is necessary.}

\section{Mechanical validation of the software synchronization of the smartphone clock}\label{sec:timecalib}
\subsection{Mechanical wave propagation}

\begin{figure}[t!]
\begin{center}
 \includegraphics[width=0.85\columnwidth]{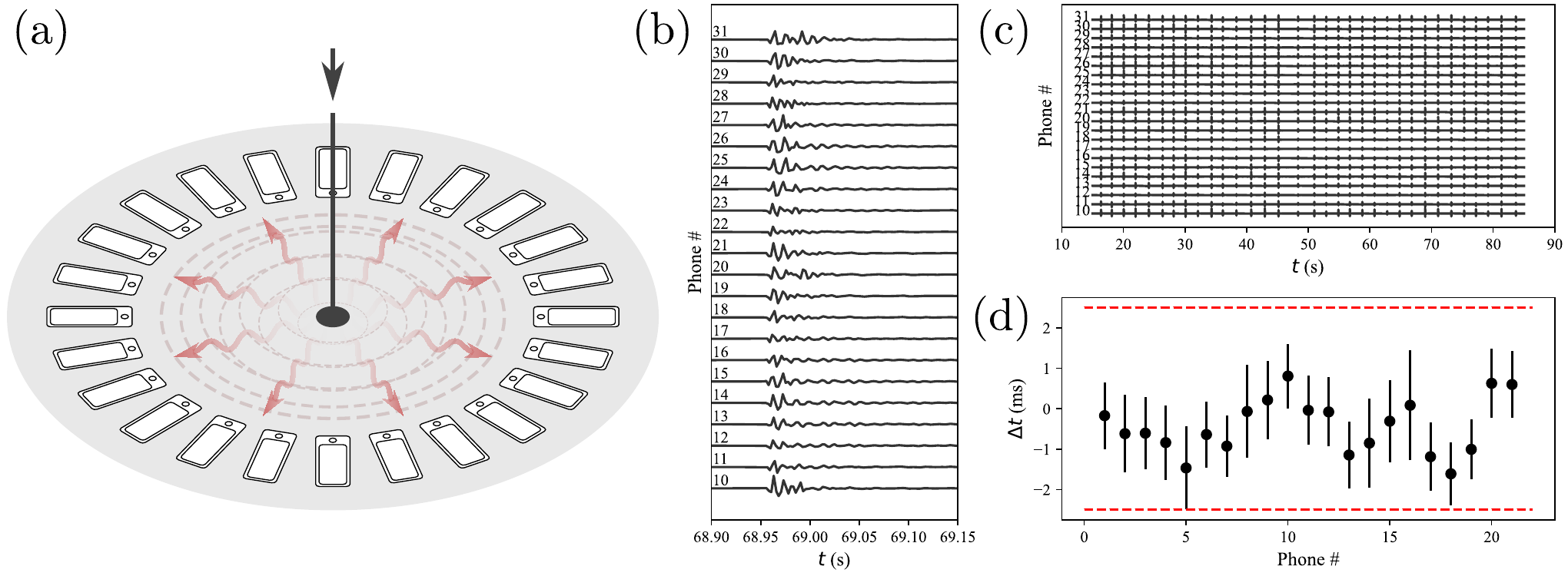}
  \caption{\label{fig:mecanicalsync} Conceptual sketch for (a) time synchronization using the propagation of Lamb waves in a stiff plate. (b) Zoom view on the vertical acceleration signal after one hit at the center of the table. The response arrives on all the phone within one sampling period. c) Display of the full recording (32 hits), showing that all phones receives each vertical acceleration within two time clock difference $2 t_s$, and smaller than one on average. d) Averaged time difference $\Delta t$ of arrival between the first phone and each of the 21 other phones. The time difference $\Delta t$ remains smaller than the sampling period $t_s = 2.5$ms for all smartphones. The errorbar corresponds to the standard deviation of the time difference distribution.}
 \end{center}
 \end{figure}

   \rev{ We have conducted experiments to mechanically test the smartphone synchronization. The acceleration signals have a sampling frequency of $400$ Hz, which implies a two-step approach to verify the validity of our value. We first placed $22$ smartphones (labeled from $\#10$ to $\#31$) on a metallic plate along a circular pattern (see fig. \ref{fig:mecanicalsync}a) and excited impulsional lamb waves from the center after synchronizing them through our custom UDP-based protocol. Fig \ref{fig:mecanicalsync}b presents such a typical impulsion as it is recorded by the 22 synchronized smartphones, showing that all smartphones record the response within one sampling period. We further confirm this by repeating a sequence of 32 hits on the plate (Fig \ref{fig:mecanicalsync}c). We can compute the averaged time difference $\Delta t$ of arrival between the first phone and each of the 21 other phones. The time difference $\Delta t$ remains smaller than the sampling period $t_s = 2.5$ms for all smartphones. The errorbar corresponds to the standard deviation of the time difference distribution. This synchronization method, while reliable, cannot provide an estimate significantly better than the sampling period $t_s$.}

\subsection{Phase measurement of two oscillating phone}

\begin{figure}[t!]
\begin{center}
 \includegraphics[width=.85\columnwidth]{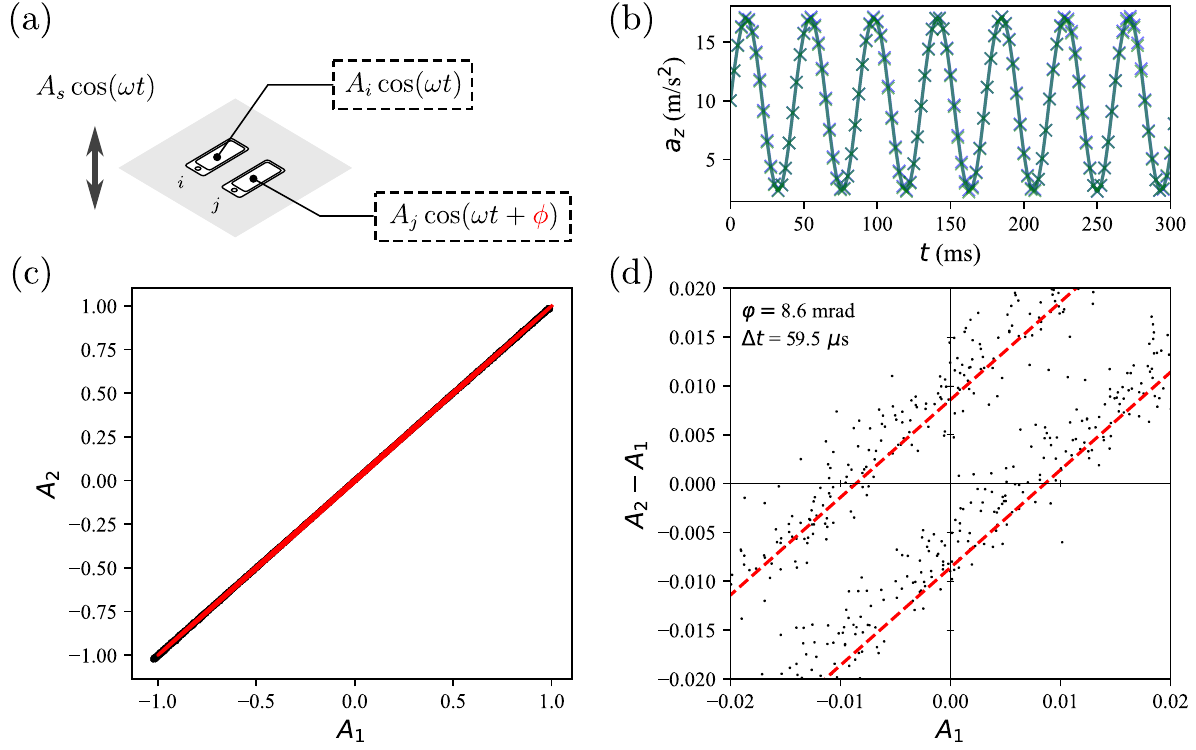}
  \caption{\label{fig:mecanicalsync_phase} Temporal signal of two phones (\# 32 \& \# 33) oscillating vertically at $f = 23$ Hz. By eye, the phase is identical. b) Normalised acceleration $A_2$ of the phone \#33 as a function of the normalised acceleration $A_1$ of phone \#32. The phase $\varphi_0$ between the two signal using $\langle A_1 A_2 \rangle_t  = 1/2 \cos(\varphi_0)$. We found $\varphi = 0.0086 \pm 0.001$ rad, corresponding to a time difference $\Delta t = 59.5 \pm 8 \mu$s between the two phones. c) Validation of the signal shape, by representing the difference $A_2-A_1$ as a function of $A_1$. The prediction of two sinusoidal signals of same amplitude with a phase difference $\varphi = 0.0086$ is superimposed in red.}
  \end{center}
 \end{figure}

\rev{To evaluate the time difference below the sampling frequency, we design a second experiment, shown in figure \ref{fig:mecanicalsync_phase}(a). We mount two phones on a vertically vibrating shaker (Bruel and Kjaerr 4808) driven by a low frequency generator providing a vertical oscillation $A=A_s \cos{\omega t}$ at $f=\omega/(2\pi) = 23$Hz. The phones are mounted tight on each other, to avoid any parasitic vibrations. The vertical acceleration $a_z$ of both phones are shown in figure \ref{fig:mecanicalsync_phase}(b). Both signal almost perfectly overlap when $a_z$ is plotted against their respective (synchronized) time. To estimate the error $\Delta t$, we perform the follwing procedure: we first plot the acceleration $A_2$ of the second smartphone against the acceleration of the first phone $A_1$ (fig. \ref{fig:mecanicalsync_phase}(c)). The signals seem to be almost perfectly in phase. To estimate the phase shift $\varphi_0$ between the two signals, we  use $\langle A_1 A_2 \rangle_t  = 1/2 \cos(\varphi_0)$. We found $\varphi = 0.0086 \pm 0.001$ rad, corresponding to a time difference $\Delta t = 59.5 \pm 8 \mu$s between the two phones. This is further confirmed when looking at the difference $A_2-A_1$ as a function of $A_1$. The prediction of two sinusoidal signals of same amplitude with a phase difference $\varphi = 0.0086$ is superimposed in red.}

\section{Operational constraints of the smartphone fleet}

This appendix summarizes practical engineering considerations encountered in our specific experimental implementation of the smartphone fleet, including power consumption, data storage, network communication, and data recovery.

\textbf{Battery consumption} Several tests were performed to evaluate battery consumption during long-duration recordings, as a function of screen state (on/off), illumination, external temperature, and smartphone protection. Both ambient temperature and screen illumination were found to significantly affect battery lifetime.

Screen state has a strong impact on battery lifetime. The most energy-efficient configuration corresponds to a screen-off setting, with short periods of activity (5min every hour), for which a periodic recording duration of 48 hours was achieved. Under identical temperature conditions, keeping the screen on reduces the maximum recording duration to 20 hours. Note that these values will strongly depend on the phone model, due to battery capability, and battery optimization softwares.

Ambient temperature also plays a critical role. When the external temperature is decreased from to -20°C, the battery lifetime of non insulated smartphone drops to approximately 8 hours. In the current configuration, the operational temperature range of a non insulated smartphone extends from $-20$ °C to $25$ °C. Finally, the use of a custom plastic protective case significantly improves low-temperature performance as their provide thermal shielding. With this enclosure, smartphones remain operational at temperatures as low as $-20$ °C, with a battery lifetime larger than 14 hours.

\textbf{Data-volume management} The Redmi 10A smartphone provides a total usable internal storage capacity of 64 GB. For continuous recordings simultaneously using the accelerometer (400 Hz), gyroscope (400 Hz) and magnetometer (50 Hz) at their maximum frame rate, we observe a data flux of 100 MB/hour. The smartphones thus run out of battery before their internal memory is full. Data volume does not represent a limiting factor in practice.

\textbf{Packet-loss rates} Packet loss was primarily observed during network communication phases, such as remote start/stop commands and data retrieval, and did not affect data acquisition itself, which is performed locally on each device. In stable network conditions, command execution and data recovery through Wi-Fi were successful for more than 95 \% of the smartphones. Packet losses were mainly associated with transient Wi-Fi instabilities, particularly in outdoor environments or during simultaneous data transfers from multiple devices. These connection ruptures were solved by manual recovery when needed, and did not result in data corruption or loss of recorded signals. Running Gobannos locally on the phone ensures that all the data are recorded on the phone memory. In this configuration, we did not experience any loss of data.

\bibliographystyle{elsarticle-num}
\bibliography{fleet}

\end{document}